\documentclass[12pt]{iopart}

\usepackage{iopams}
\usepackage{graphicx}
\begin{document}

\title[Investigation of Energy Spectrum of EGRET Gamma-ray...]{Investigation of Energy Spectrum
of EGRET Gamma-ray Sources by an Extensive Air Shower Experiment}

\author{M. Khakian Ghomi$^{2}$, M. Bahmanabadi$^{1,2}$, F. Sheidaei${^1}$, J. Samimi$^{1,2}$, A. Anvari$^{1,2}$}

\address{$^1 $Department of physics, Sharif University of Technology,\\
 P.O. box 11365 - 9161, Tehran, Iran.\\
$^2$ALBORZ Observatory(http://sina.sharif.edu/$^\sim$observatory)}
 \ead{sheydaei@sharif.edu}

\section*{abstract}
Ultra-High-Energy (UHE) ($E>100~$TeV) Extensive Air
Showers (EASs) have been monitored for a period of five years
($1997-2003$), using a small array of scintillation detectors in
Tehran, Iran. The data have been analyzed to take in to account of
the dependence of source counts on zenith angle. Because of varying
thickness of the overlaying atmosphere, the shower count rate is
extremely dependent on zenith angle. During a calendar year
different sources come in the field of view of the array at varying
zenith angles and have different effective observation time
equivalent to zenith in a day. High energy gamma-ray sources from
the EGRET third catalogue where observed and the data were analyzed
using an excess method. Upper limits were obtained for 10 EGRET
sources \cite{khakian}. Then we investigated the EAS event rates for
these 10 sources and obtained a flux for each of them using
parameters of our experiment results and simulations. Finally we
investigated the gamma-ray spectrum in the UHE range using these
fluxes with reported fluxes of the EGRET sources. \\
\textbf{keywords:}{The EGRET sources, Extensive Air Showers (EASs),
Gamma-ray sources, Gamma-ray spectrum}

\maketitle
\section{Introduction}
 The EGRET instrument on-board  Compton gamma-ray Observatory (CGRO)
has detected about 271 high energy ($>100~$MeV) gamma-ray sources
\cite{hartman}. Effective sensitivity of EGRET is in the energy
range from 100~MeV to 30~GeV.\\ The EGRET gamma-ray sources in many
aspects like characteristics, different energy ranges and etc. have
been investigated \cite{Bhattacharia,cillis,zhang}. Whether the
EGRET sources emit gamma-ray at still higher energies or not, is an
interesting question. Gamma-rays with energies of about 100 TeV and
more, entering the earth atmosphere, produce Extensive Air Shower
(EAS) events which could be observed by the detection of the
secondary particles of the EAS events on the ground level
\cite{gaisser}. This gamma-ray induced EAS events are investigated
for diffuse Galactic \cite{Brezin} and extragalactic sources, also
Galactic \cite{atkins}, \cite{mac} and extragalactic \cite{fidelis}
gamma-ray point sources have been investigated too. In this work we
have investigated 10 of these point sources.\\ Our small particle
detector array is located at the Sharif University of Technology in
Tehran, Iran at about 1200~m above sea level
($\equiv$~890~g~cm$^{-2}$). This small array is a prototype for a
large EAS array to be built at an altitude of 2600~m
($\equiv$~756~g~cm$^{-2}$) at ALBORZ Observatory (AstrophysicaL
oBservatory for cOsmic Radiation on alborZ) (http://sina.sharif.edu/
$^\sim$observatory/) near Tehran. The results of the experiments
with this prototype observatory with about $1.7\times 10^{5}$
recorded EAS events was reported earlier \cite{khakian} and has been
shown that some of the EGRET GeV point sources are gamma-ray
emitters at energies about 100TeV. Here we report the results of our
recent investigation using the data of our earlier experiments to
extend the energy spectrum of the observed EGRET sources up to
100TeV and more. In this investigation we present the observational
results for 10 EGRET third catalogue sources. Then we investigate
the effective area and effective time of observation of each source.
Finally we compare the obtained fluxes and spectral indices with the
presented fluxes and spectral indices of the EGRET third catalogue
sources at the 3rd EGRET catalogue.
\section{Experimental arrangement}

Our array is constructed of four slab of plastic scintillation
detectors ($100\times100\times2$~cm$^3$). They are housed in white
painted pyramidal boxes \cite{PhDbahmanabadi} which arranged in a
square; at 51$^{\circ}$~20$^{'}$E and 35$^{\circ}$~43$^{'}$N,
elevation 1200~m ($\equiv$~890~g~cm$^{-2}$). Two different
experimental configurations were used in the experimental set up.
The first ($E1$) and the second ($E2$) experimental configurations
are identical except the size of the square array. In $E1$ the size
is 8.75~m~$\times$~8.75~m and in $E2$ the size is
$11.30$~m~$\times$~$11.30$~m. More details of the experimental setup
is given in the reference \cite{khakian}.
\section{Data Analysis}

The logged time lags between the scintillation detectors and
Greenwich Mean Time (GMT) of each EAS event were recorded as raw
data. We synchronized our computer to GMT
(http://www.timeanddate.com). Our electronic system has a recording
capability of 18.2 times per second. If an EAS event occurs, its
three time lags will be recorded and if it does not occur 'zero'
will be recorded. Therefore the starting time of each experiment and
the count of records gives us the GMT of each EAS event. We
estimated the energy threshold and the mean energy of our experiment
and also we calculated the statistical significance of 98 of the 3rd
EGRET sources which were in the Field Of View (FOV) of our array.
\\
The complete analysis procedure \cite{khakian} is itemized as
follows:
\begin{itemize}
 \item The local coordinates: zenith and azimuth angles of each
 EAS event $(z,\varphi)$ were calculated using a least-square method
 based on the logged time lags and coordinates of the scintillators.
 A zenith angle cut off of $60^{\circ}$ is implemented to increase the
 significance \cite{mitsui}.
 \item The distributions of the local angles of the EAS events were
 investigated to understand the general behavior of these events.
 We fitted these distributions with the two functions as follows \cite{magnetic}:

 \begin{equation}
 \\dN=A_{z}\sin z\cos^n z dz
 \end{equation}
 where $A_z=95358$ and $n=5.00$, and also
 \begin{equation}
 \\f(\varphi)=A_{\varphi}+B_{\varphi}\cos(\varphi-\varphi_1)+C_{\varphi}\cos(2\varphi-\varphi_2)
 \end{equation}
 where $A_{\varphi}=1,B_{\varphi}=0.081 $ , $C_{\varphi}=0.069$, $\varphi_{1}=95^\circ$ and $\varphi_{2}=-193^\circ  $.
 \item Celestial coordinates (RA,Dec) of each EAS event were
 calculated
 using its local coordinates, the GMT of the event and geographical
 latitude of our array (http://tycho.usno.navy.mil/sidereal.html).
 Then we calculated galactic coordinates (\emph{l},\emph{b}) of each EAS event
 from its equatorial coordinates for epoch J2000 \cite{osouloamal}.
 \item We estimated the errors in (\emph{l},\emph{b}) of the investigated EGRET sources
 from the error factors in the array. In this stage we obtained
 $\bar{r_e}=4.35^{\circ}\pm0.82^{\circ}$ as the mean angular error
 of our experiment.
 \item We investigated cosmic-ray initiated EAS events by simulations based on a homogeneous
 distribution of primary charged particles. These simulations incorporated all known parameters of the
 experiment. \cite{khakian}
 \item We investigated the statistical significance of 98000 random sources and
 also 98 sources of the 3rd EGRET catalogue, using the method
 of Li \& Ma (Li \& Ma 1983)
 we derived the best-known locations for the EGRET
 sources in the TeV range \cite{khakian}.
 \end{itemize}
\subsection{Shadow of the moon}
Observing the shadow of the moon in EAS experiments which usually
might have a much larger error circle than the disk of the moon is a
very difficult task and requires a careful scrutinization of the
data. The difficulty is compounded since a realistic radius for the
error circle of the experiment is only obtained by the observation
of the shadow which could be indicated as a deficit of shower counts
falling in the error circle centered about the moving location of
the moon as compared to the average shower counts falling in error
circles centered at other positions in sky during the observation
time. To carry out the scrutinization of our data, we have proceeded
as follows.We have divide our data into sequential time segments and
for each time segment we have used the mean values of the local
coordinates of moon $(\theta_{m},\varphi_{m})$ moving over our
observatory site. These coordinates were obtained using the
information provided by the http://aa.usno.navy.mil. Now for an
assumed radius of circle of error, angular radius $\rho_{err}$
ranging from $0.5^{\circ}$ to $ 15^{\circ}$ incremented by
$0.5^{\circ}$, we have calculated the number of showers falling in
each circle. This has been done by calculating the angular
separations $\Theta_{s}$ between the arrival direction of each
shower event $(\theta_{s},\varphi_{s})$ and the direction of the
moon at the time of recording of that event, using the following
equation from spherical geometry:
 \begin{equation}
\cos\Theta_s=\cos\theta_{m}\cos\theta_{s}+\sin\theta_{m}\sin\theta_{s}\cos(\varphi_{m}-\varphi_{s})
 \end{equation}

Obviously if $\Theta_{s}<\rho_{err}$ that shower is counted as
falling in the moon's error circle. In order to compare the obtained
result with random sampling and scrutinize the difference for each
value of $\rho_{err}$ we have chosen 1000 random locations in the
sky denoted by local coordinates $(\theta_{r},\varphi_{r})$ and have
calculated the number of showers falling in the error circles
countered about each of the 1000 random locations. This was
similarly done by calculating the angular separation of each shower
arrival direction $(\theta_{s},\phi_{s})$ with the direction of the
center of the randomly chosen error circle $(\Theta_{r})$ from above
equation with $(\theta_{m},\varphi_{m})$ replaced by
$(\theta_{r},\varphi_{r})$. If for any shower event
$\Theta_{r}<\rho_{err}$ that shower is counted as falling in the
error circle of that random position. In these computations (both
for the moon as well as for the random locations) a weight factor
was used for each of the shower events in order to account for the
site-specific effects in our data which depend on the arrival
directions of shower events. These effects are :\\
(1) The different thickness and density of the overlying atmosphere
which are effective in shower development and hence its
detectability at the height of the observatory.\\
 (2) The geomagnetic effect on the azimuthal arrival direction of the
showers. These effects which are specific for each observation site,
were separately and independently determined for our site and are
reflected in the dependence of the number of shower events on
zenithal and azimuthal angles which are given by equations 1 and 2,
respectively \cite{magnetic}. To take account of these effects in
our observed data, we have assigned a weight factor to each shower
arrival direction which is the product of these two independent
factors. The weight factor is thus:
\begin{equation}
W(\theta_{s},\varphi_{s})=\cos^{n}\theta_{s}(A_{\varphi}+B_{\varphi}\cos(\varphi-\varphi_1)+C_{\varphi}\cos(2\varphi-\varphi_2))
\end{equation}
 where the constants n, $A_{\varphi}$,$B_{\varphi}$ and $C_{\varphi}$ are
given below Eqs.\textbf{1} and \textbf{2}.\\
 It should be remembered that in our earlier
work, \cite{khakian} reporting the observation of EGRET gamma-ray
point sources in TeV data by excess method, both of these effects
were carefully taken into account by determining the exposure map of
our observations by simulations incorporating these effects along
with other particularities of our observations and by correcting of
our observed data by dividing it by the exposure map in galactic
coordinates.

\subsection{Distributions of number of EAS events in error circles}
Here we discuss the distribution of number of EAS events falling in
the error circles of different radii as determined according to the
aforementioned procedure. We first present the results for each of
the sets of 1000 randomly chosen error circles. Fig.2 shows the
histogram of frequency of occurrence of number of circles with
respect to the number of EAS falling in the error circle. The
histograms distributions corresponding to different radii of error
circles are shown here as illustration. These for other radii show
similar distributions and all these distributions nearly fit
gaussian distributions with a mean and a variance which increase
with increasing radius. This result is very assuring and shows the
correctness of our sophisticated numerical procedure and the
validity of using the mean number of events, in these randomly
chosen locations of error circles in the sky to compare with the
number of events falling in the error circle countered about the
moving moon. Fig.3 shows the variation of the mean number of events
of these distribution as a function of the chosen radius of the
error circles. It is seen that these calculated means show a nearly
exact dependence on the square of the radius. This result is not
surprising and is exactly what one would expect to get from a
correct random sampling of statistical data. However, in view of the
complexity and sophistication involved in our entire procedure (with
inclusion of the weight factors), this result is again very assuring
and shows that we can use these mean number of events to compare
with that falling in the error circles centered about the moving
moon and rule out the possibility of the deficit in the number of
events falling in the moon centered circle as due to statistical
fluctuation. In Fig.3 we have also shown the number of events
falling in the moon-countered circles for comparison with the means
of the random samplings. It is seen that in every case( for every
chosen error circle radius) the deficit of number of EAS from the
direction of the moving moon is quite significant as compared to the
error bars of the mean of random sampling( Fig.3) which is taken as
the variance of nearly gaussian distributions of Fig.2. As seen here
the deficit which is from 1.6 to about 7.1 times the standard
deviation of the mean of random distributions is quite significant
and since it is definitely associated with the moving moon it must
be associated with some moon- related phenomenon. We will not
discuss this phenomenon here and rather simply call it the shadow of
the moon in our EAS observed data.
\subsection{Estimation of energy thresholds}

Our detected EAS events are a mixture of cosmic-ray and gamma-ray
events. In $E1$ the total number of EAS events was 53,907 and the
duration of the experiment was 501,460 seconds. So the mean event
rate of the first experiment was 0.1075 events per second. The
distribution of the time between successive events was investigated
and found to be in good agreement with an exponential function,
indicating that our event sampling is completely random
\cite{anisotropy}. In $E2$ the total number of events was 173,765
and the duration of the second experiment was 2,902,857 seconds, so
its mean event rate was 0.05986 events
per second.\\
We refined the data to separate out the acceptable events. Events
are acceptable if there is a good coincidence between the four
scintillator pulses, also we omitted the events with zenith angles
more than 60$^{\circ}$ because of their less accuracy. Therefore
after these separations we obtained smaller data sets of 46,334 and
120,331 events for $E1$ and $E2$ respectively. Since we cannot
determine the energy of the showers on an event by event basis, we
estimate our lower energy threshold by comparing our event rate to
the following cosmic-ray integral spectrum \cite{Borione},
\begin{equation*}
J(E) = 2.78\times10^{-5}E^{-2.22} + 9.66\times10^{-6}E^{-1.62}
\end{equation*}
\begin{equation}
\hspace{1cm}-1.94\times10^{-12} \hspace{3mm} 40\leq E \leq 5000
\hspace{1mm}{\rm TeV}
\end{equation}

The obtained lower energy limits are 39 TeV in $E1$ and 54 TeV in
$E2$. The calculated mean energies with above energy spectrum are 94
and 132 TeV in $E1$ and $E2$, respectively. Since the distribution
of cosmic-ray events within the array around these energy ranges is
homogeneous and isotropic, we used an excess method \cite{tibet} to
find signatures of the EGRET 3rd catalogue gamma-ray sources. This
method was used for both $E1$ and $E2$.
\section{Calculation of effective area and time}
Number of secondary particles in the growth profile of EAS events
increases in atmosphere until the shower maximum and then decreases
after it. In energy of about 100~TeV the shower maximum height is at
about 500g~cm$^{-2}$ and a fraction of these secondary particles
arrive to the ground level particle detectors of our array at Tehran
level (1200~m$\equiv$ 890 g~cm$^{-2}$).\\ For calculating the
effective surface of each experiment ($E1$ and $E2$) we used Greisen
lateral distribution of electrons which is known the NKG formula
\cite{kamata}
 and CORSIKA simulation code
\cite{heck} for the simulation of the two sets of proton showers
with energy thresholds of 39~TeV and 54~TeV respectively for $E1$
and $E2$  at our array level. From our logged EAS events we obtained
a zenith distribution function for $E1$ and $E2$, which the mean
zenith angles are $26^{\circ}$ for both of them. Also we calculated
the $\bar{z}$ which was obtained from the weight curve of the zenith
distribution $dN/dz\propto\cos^nz\sin{z}$ and we obtained the
$26^{\circ}$ too. This weight curve was obtained by fitting the
function $dN/dz$ to our data in the $E1$ and $E2$ with $n=5.00$. So
in the first approximation we used the effective thickness of the
passed atmosphere as $890{\rm g~cm}^{-2}/\cos(26^{\circ})= 980{\rm
g~cm}^{-2}$. In the thickness, the average number of the secondary
particles for the two experiments are $N_{E1}=5265$ and
$N_{E2}=8571$, these two numbers obtained from 1000 simulated proton
showers for each energy threshold. So based on the NKG formula the
mean effective surfaces of EAS events at Tehran level are $965~{\rm
m}^2$ and $2173~{\rm m}^2$ for $E1$ and $E2$ respectively. With
these results we could obtain the mean effective surface of our
array in the upper level of the atmosphere (The surface that if a
primary particle passes through it, the array could detect its EAS
events) $718{\rm m}^2$ and $1751{\rm m}^2$ for $E1$ and $E2$
respectively.\\ For calculation of the effective time of observation
of each source in every 24 hours, we used the spherical geometry and
the track of each source in the local coordinates. Each source with
its celestial coordinates right Ascention, Declination (RA,Dec) is
introduced in the 3rd EGRET catalogue. Time duration of each source
is calculated by reaching the source to the zenith angle of
$60^{\circ}$ from the direction of east to the same zenith angle
from the west, and the distribution function
$dN/dz=\propto\cos^nz\sin{z}$ which is related to the zenith
distribution effect \cite{khakian}. Finally we obtained the mean
effective time of observation of our array for all 10 sources
equivalent to 4h,28$^{'}$ (equivalent to existence of 4$h$,28$^{'}$
the source at zenith) for every 24 hours. The FOV of our array with
the $60^{\circ}$ zenith angle cutoff is $\pi$ steradian. With these
calculated factors we obtained fluxes(events
cm$^{-2}$s$^{-1}$sr$^{-1}$) for each of the 10 sources in $E1$ and
$E2$ which are shown in Table 1.
\section{Results}
Our results have been compared with the EGRET results. For each
source we have fluxes and energies from EGRET, $E1$ and $E2$, so we
extracted a spectral index for each source and compared it with the
reported spectral index of EGRET. Some information about the 10
EGRET sources like Name, RA, Dec, and spectral index and its error,
$(\gamma\pm\Delta\gamma)_{EGT}$, are from the 3rd EGRET catalogue
\cite{hartman}. Other information like mean energy for $E1$ and $E2$
which are 94TeV and 132TeV respectively (are not shown in the
table), fluxes of the two experiments and spectral indexes and their
errors, $(\gamma\pm\Delta\gamma)_{OUR}$, have been calculated in
this analysis. The last column shows the agreement of our spectral
indices to its in the 3rd EGRET catalogue.

\subsection{Result of Energy Analysis with Simulated Showers}
For further energy analysis of our measured data of EAS events
observed at our site we have used the CORSIKA code\cite{heck} to
simulate showers with the inclusion of geomagnetic field pertinent
to the location of our site from data provided by
http://www.ngdc.noaa.gov. We have simulated a total of 7350 showers
entering the top of atmosphere at various zenith angles
$(0^{\circ},15^{\circ},30^{\circ},45^{\circ},60^{\circ})$ and for
each zenith angle at 12 various azimuth angles ranging from
$0^\circ$ to $360^{\circ}$ every $30^{\circ}$. For each angle the
simulations were repeated 10 times for energies less than 50 TeV and
5 times for energies greater than 50 TeV with separately 10 TeV
intervals from 10 TeV to 100 TeV. These simulations were carried out
similarly for entering protons as well as gamma-rays. For each
simulation the number of secondary charged particles of the
simulated shower at the height of observation of our site was
determined from CORSIKA code. In order to make an energy analysis
comparing gamma-ray initiated showers with those initiated by
protons in our observations we have calculated a mean number of
secondary shower particles for each energy by averaging over all the
angles using the site-specific weight factor discussed in sec
3.1(Eq. 4). The ratio of angle-averaged mean of number of secondary
charged particles in gamma-ray initiated showers to that of the mean
for proton-initiated showers is shown in Fig.4 as a function of
energy in the energy range of our simulations. The ratio shown here
incorporates almost all of related particularities of our
experiments. Since in our observations we have used exactly
identical experimental procedure, equipment, thresholds and set-ups
for all of the recorded showers irrespective of the nature of the
radiation which initiates the shower the relative detection
efficiency of our array and observations could only depend on the
ratio of these means depicted in Fig.4. Now we use the relative
detection efficiency calculated from simulated data and shown in
Fig.4 to estimate the relative number of gamma-ray initiated showers
to that initiated by protons, using the known energy spectrum of
proton showers of the form
\begin{equation}
\frac{dN_{p}}{dE}=N_{p10}(\frac{E}{10TeV})^{-\Gamma_p}.
\end{equation}
Where $dN_{p}$ is the number of proton showers in the energy
interval $E$ to $E+dE$, $N_{p10}$ is a constant and $\Gamma_p$ is
the spectral index of EAS producing protons. Now denoting our
array's relative detection efficiency by $\eta(E)$, we can write for
the expected differential number of gamma-ray initiated showers:
\begin{equation}
\frac{dN_{\gamma}}{dE}=\eta(E)
N_{p10}(\frac{E}{10TeV})^{-\Gamma\gamma}
\end{equation}
where $\Gamma_\gamma$ is the spectral index of EAS producing gamma-
rays. Dividing Eq.7 by Eq.6 and upon integration from a threshold
energy to infinity, the ratio of observed gamma-ray EAS events to
that initiated by protons is obtained. Thus we write:\\

\begin{equation}
\frac{N_{\gamma EAS}}{N_{pEAS}}=\frac{\int^{\infty}_{E_{t}} \eta(E)
E^{-\Gamma_\gamma}\, dE}{\int^{\infty}_{E_{t}}  E^{-\Gamma_{p}}\,
dE}
\end{equation}
where $E_t$ is assumed threshold energy. For the energy range
covered in this analysis, we use $\Gamma_{p}=2.7$ and for gamma-ray
spectral index we use the mean value of the spectral indices that we
have estimated above (Table.1) for the sources with the highest
statistical significance (those observed at a statistical
significance level of higher than $1.5\sigma$). Thus using our
estimations of table 1 we have $\Gamma_{\gamma}=2.3$. In order to
carry out the integrations in Eq.8 we are forced to impose a
truncation since with our rather limited computer shower simulations
we only have $\eta(E)$ for $10 TeV<E<100 TeV$. For the sake of
numerical consistency, we have imposed this truncation to E=100 TeV
in the denominator of Eq.8  as well as its numerator. The result of
these calculations obviously depend on the value of an assumed
threshold energy, $E_{t}$, therefore the numerical calculations were
repeated for values of $E_{t}$ from 10 TeV to 100 TeV with 10 TeV
intervals. The result of these calculations, that is, the ratio of
the number of gamma-ray EAS events to that of proton EAS events
expected to be detected in our experiments at our site is shown as a
function of the threshold energy in Fig.5. Here we see that this
ratio shows a maximum at a value of $E_{t}=40\,TeV$ which very
nearly corresponds to the lower value of the two threshold energies
of our two experiments.
\begin{table}
  \begin{tabular}{ccccccccccc}
\hline\hline
    Name($3EG\hspace{1mm}J$)&RA,Dec&$ID$&$(\gamma\pm\Delta\gamma)_{EGT}$&$\log{F_{E1}}$&$\log{F_{E2}}$&
    $(\gamma\pm\Delta\gamma)_{OUR}$&$|\gamma_{OUR}-\gamma_{EGT}|$\\
    \hline
    0237+1635&39.3,16.5&A&1.85$\pm$0.12&-11.25&-11.53&1.90$\pm$0.27&0.05\\
    0407+1710&61.8,17.1& &2.93$\pm$0.37&-10.77&-11.45&4.20$\pm$0.40&1.27\\
    0426+1333&66.6,13.5& &2.17$\pm$0.25&-11.11&-11.28&1.15$\pm$0.48&1.02\\
    0808+5114&122.1,51.2&a&2.76$\pm$0.34&-11.01&-11.58&3.87$\pm$0.42&1.11\\
    1104+3809&126.1,38.1&A&1.57$\pm$0.15&-11.10&-11.72&4.21$\pm$0.50&2.64\\
    1308+8744&197.0,87.7& &2.17$\pm$0.66&-10.96&-11.53&3.87$\pm$0.45&1.70\\
    1608+1055&242.1,10.9&A&2.63$\pm$0.24&-10.92&-11.72&4.34$\pm$0.50&1.71\\
    1824+3441&276.2,34.6& &2.03$\pm$0.50&-10.65&-11.25&4.07$\pm$0.35&2.04\\
    2036+1132&309.1,11.5&A&2.83$\pm$0.26&-11.08&-11.73&4.41$\pm$0.53&1.58\\
    2209+2401&332.4,24.0&A&2.48$\pm$0.50&-10.91&-11.31&2.71$\pm$0.42&0.23\\
  \end{tabular}
  \caption{Comparison of our spectral indices of the 10 sources with the spectral indices
  of them which is introduced by the 3rd EGRET catalogue. The last column shows agreement
  of our spectral indices to EGRET spectral indices.}
\end{table}
\section{Concluding remarks}
We believe the main reasons for success in these observations and
investigations on EGRET gamma-ray point sources despite the low
statistics of $\sim3\times10^{5}$ EAS events from our small array of
ALBORZ prototype observatory, relies on the following two favorite
points of strength:\\ (\textbf{1}) we have intensively studied the
location- dependent factors which influence shower development and
shower count from various angular bins in the sky. These factors are
the mass of the overlying atmosphere and anisotropy in azimuth
angles which it is attributed to the effect of the geomagnetic field
\cite{Ivanov}, \cite{magnetic}, \cite{He}. For our particular site
we investigated these effects which are in form of Eqs. 1 and 2. We
carefully refined our data from these effects finally the
investigation of the EGRET gamma-ray point sources are based on the
corrected data.
\\(\textbf{2}) Our site and the duration of our observations in the
data set have been such that 10 of the EGRET gamma-ray point sources
have crossed our site at small enough zenith angles to make their
observation possible at least with a statistical significance of
$1.5\sigma$ with Li and Ma criterion in the excess method which we
have used\cite{khakian}. In order to investigate this rather
fortunate point for our site and the time of these observations, we
have calculated the average zenith all angle of passage of each of
271 EGRET gamma-ray point sources during our entire observation
period, for each of these sources. We have also calculated the
statistical significance(number of sigma) for excess counts from
these sources. The result is shown here in Fig.1 and it convincingly
shows an inverse correlation between the statistical significance of
excess shower counts from the EGRET point source location and the
average zenith angle of the transit of the source over our site. The
computed coefficient of this correlation(correlation between inverse
statistical significance and $\overline{z}$)  is 0.77.\\
Obviously the EGRET sources investigated here are those with highest
statistical significance corresponding to the
lowest zenith angles of the passage of the source over our site.\\
Our results shows that our spectral indices within error limits are
in agreement with EGRET  spectral indices for most of the considered
here.\\ For the calculation of the effective area of the EAS events
at first , we calculated the primary particle energy, then we
extracted the number of the secondary particles from a set of
CORSIKA simulations, and finally we used the Greizen lateral
distribution for these particles. In the near future with a larger
array and larger set of logged EAS events at ALBORZ observatory we
hope to calculate
these results more accurately.\\
It is worth remarking that due to increased probability of pair-
production interaction of VHE and UHE gamma-rays with various low
energy universal photons, the absorption of these gamma-rays becomes
significant for sources located at cosmological distances. This
absorption has been studied extensively
\cite{stecker,Nature,djager}. The lack of observation of VHE and UHE
gamma-rays from extragalactic EGRET gamma-ray sources has been
generally attributed to the absorption of the gamma-rays from these
sources which are located at
cosmological distances.\\
Stecker and De Jager (1997) gave a parametric relation for optical
depth $(\tau_{j})$ as a function of the source gamma energy
$(E_{TeV})$ and red shift (z) for the two models (j=1,2) they have
studied. For the 39TeV energy threshold of our experiments, we use
their parametric relation with an optical depth of unity to
calculate the red shift of the sources we have observed in our EAS
experiments and have investigated here(Table 1). The result is
z=0.0032 for their model 1 and z=0.0024 for their model 2.
Ong,\cite{ong} have listed the red shifts of two classes of EGRET
extragalactic sources and the calculated red shifts for the sources
investigated here is smaller than red shifts listed for these
classes of objects. However, we should remark that the recent
discovery of the unexpected by hard spectra of the Blazar sources
observed in HESS data \cite{hess} bear as the possibility of
observation of EGRET gamma-ray point sources at higher energies.
It would be open for future ground-base observations. \\
 The energy analysis carried out here using simulated shows
with CORSIKA has shown that the ratio of number of gamma-ray
generated EAS events expected to be detected at our site relative to
the number of proton- generated EAS events show a maximum at a value
of 40 TeV for the threshold energy of observation of EAS events at
our site. The fortunate fact that this value corresponds to the
lower value of the threshold energies of our two experiments,
provides further support for the fact that the data collected in our
experiments at our site has had this extra advantageous attribute
for the observation of gamma-ray initiated EAS events in the 10 TeV
to 100 TeV energy range and thus the extra advantage for observing
gamma-ray point sources in this range.
\section{acknowledgements}
This research was supported by a grant from the national research
council of Iran for basic sciences. The many useful and conductive
comments by anonymous referee is very much appreciated. Also many
thank from Prof. James Matthews for his comments and attentions to
our works at ICRC 2005, Pune, India.\\


\newpage
\begin{figure}
\includegraphics[width=14cm]{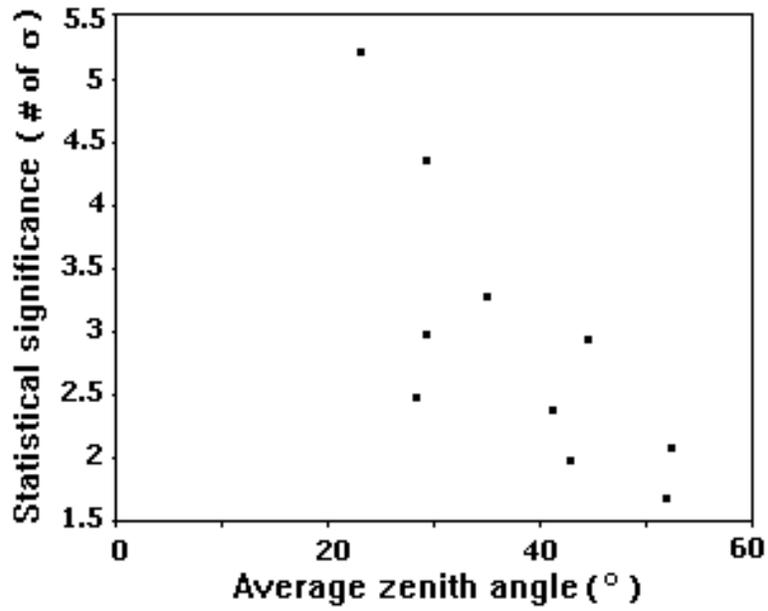}
      \caption{Distribution of statistical significance versus average zenith angle of transit of calculated for 271 EGRET gamma-ray sources over our site(only sources with significance $>1.5\sigma$ are shown.).}
         \label{setup}
   \end{figure}
       \begin{figure}
   \centering
   \includegraphics[height=10cm,width=15cm]{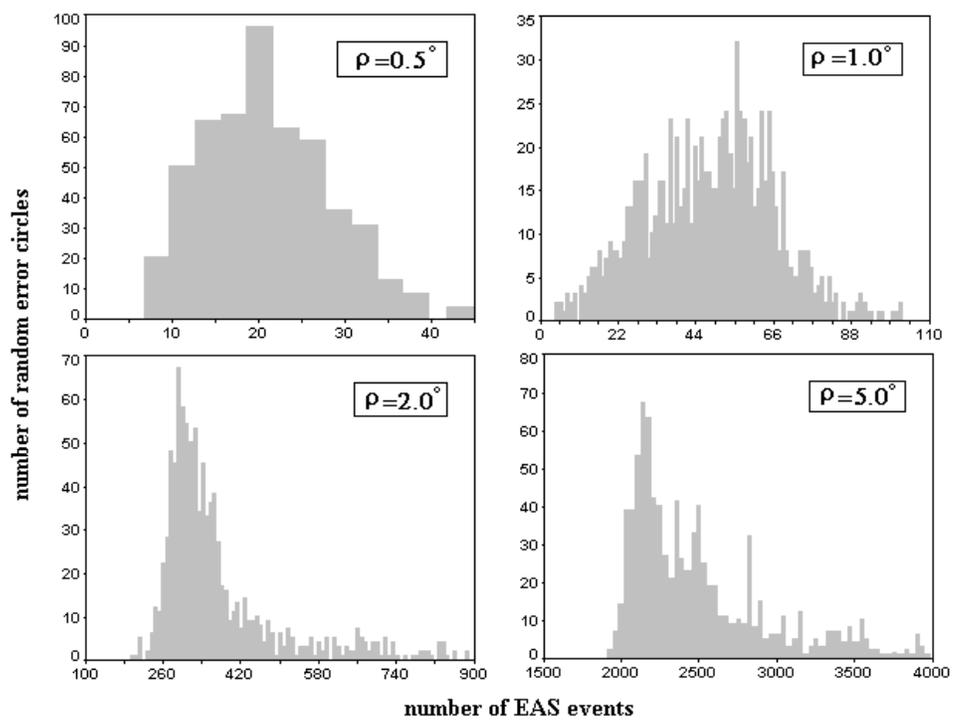}
      \caption{Histograms show the frequency of circles of error with radius $\rho$ and random chosen centers in the sky vs. the
      number of EAS events falling in each error circle.}
         \label{setup}
   \end{figure}
    \begin{figure}
   \centering
   \includegraphics[height=10cm,width=15cm]{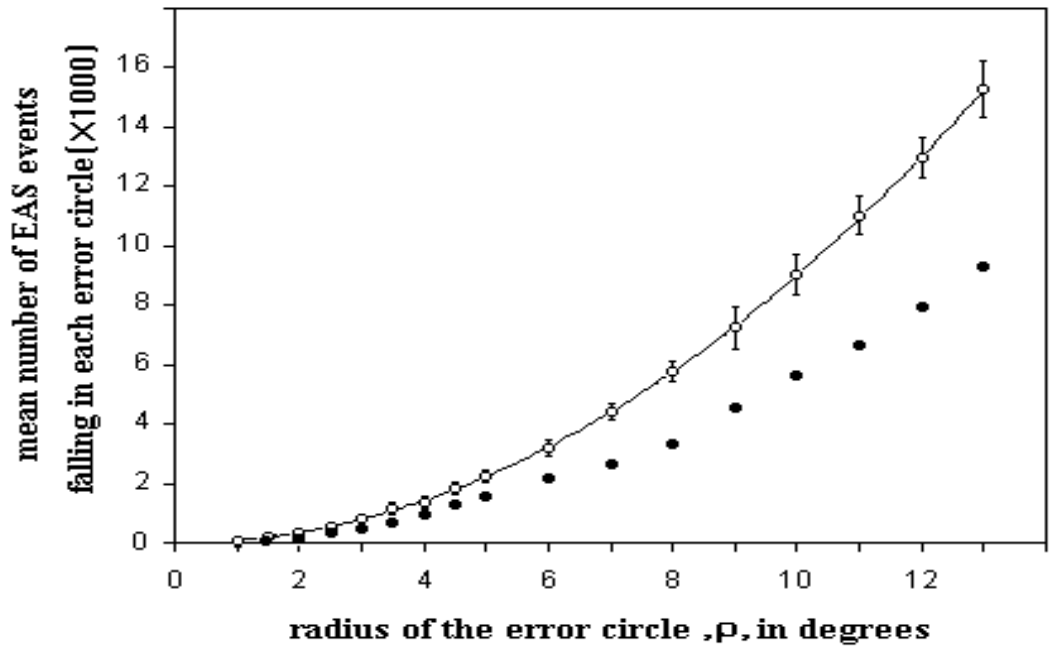}
      \caption{Dependence of the mean number of EAS events falling in the random error
       circles as a function of the radius of the error circle, $\rho$.
       Smooth curve shows a $\rho^2$ dependence. Error bars are the standard deviation
       of the distribution shown in Fig.2, and statistical
      errors for the moon. Open circles
       $(\circ)$ are random error circles. Filled circles$(\bullet)$ are error circles centered about
the location of
      the moving moon.}
         \label{setup}
   \end{figure}
  \begin{figure}
   \centering
   \includegraphics[height=10cm,width=15cm]{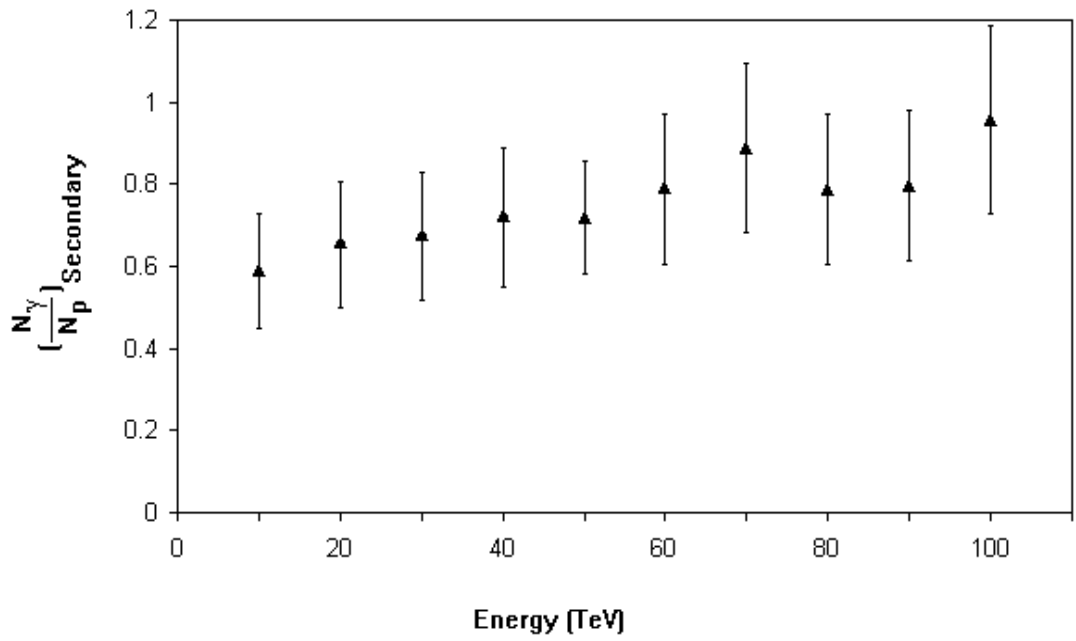}
      \caption{The ratio of the mean number of secondary charged particles at the location of our site
      produced in gamma-ray generated simulated Extensive Air Shower to that produced by proton generated showers as a function of the energy of the primary
      radiation( Gamma-ray and proton) entering the top of the atmosphere.}
         \label{setup}
   \end{figure}
 \begin{figure}
   \centering
   \includegraphics[height=10cm,width=15cm]{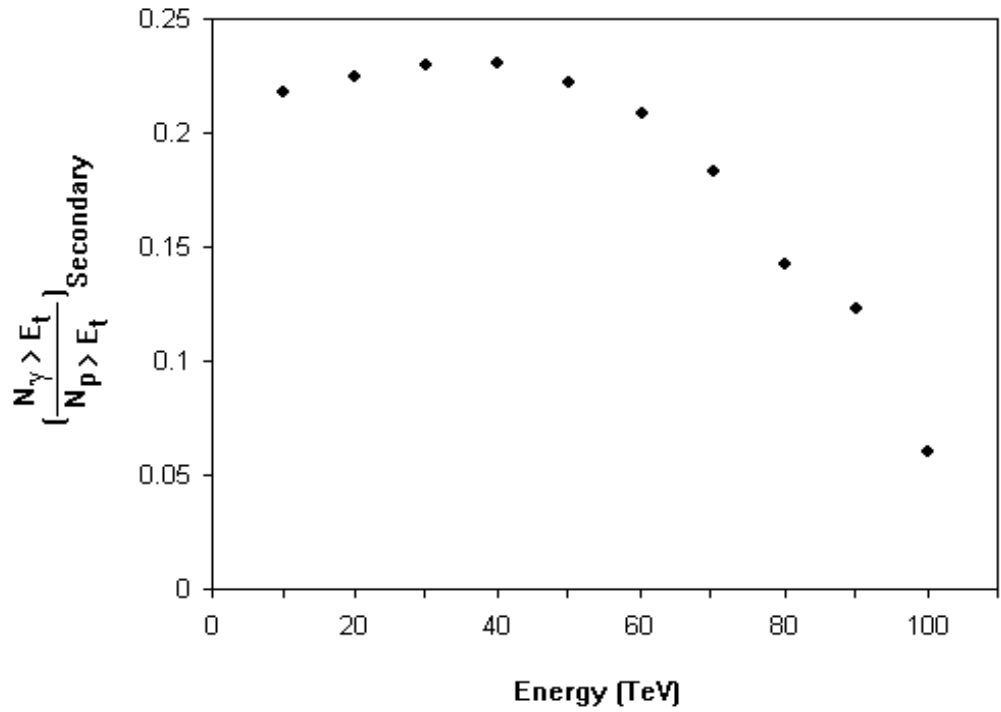}
      \caption{The ratio of the mean number of gamma-ray generated showers to that of proton generated
      showers expected to be observed at our site as a function of the threshold energy of our experiments.}
         \label{setup}
   \end{figure}
   \end{document}